\documentclass[10pt]{article}
\usepackage[letterpaper,left=2in,right=0.8in,top=0.9in,bottom=1in]{geometry}
\usepackage[T1]{fontenc}
\usepackage{newtxtext,newtxmath}%
\usepackage{booktabs,amsmath,adjustbox,graphicx,url,xcolor}
\usepackage[authoryear,round]{natbib}
\usepackage{titlesec}
\usepackage[labelfont={bf,sf},labelsep=period,font=small]{caption}
\usepackage{fancyhdr,lastpage}
\usepackage[hidelinks]{hyperref}

\graphicspath{{rysunki/}}

\titleformat{\section}{\sffamily\bfseries\large}{}{0pt}{\MakeUppercase}
\titlespacing*{\section}{0pt}{12pt}{4pt}
\titleformat{\subsection}{\sffamily\bfseries\normalsize}{}{0pt}{}
\titlespacing*{\subsection}{0pt}{8pt}{2pt}
\titleformat{\paragraph}[runin]{\sffamily\bfseries\small}{}{0pt}{}
\titlespacing*{\paragraph}{0pt}{6pt}{0.6em}
\definecolor{abstractbg}{RGB}{186,197,245}
\begin{document}
\thispagestyle{fancy}

{\raggedright\sffamily\bfseries\LARGE Are Calibrated Decision Models Calibrated on Markets? Geolocated News, Markov
Switching and a System One Model for Nasdaq-100 Allocation\par}
\vspace{10pt}
{\sffamily\bfseries\large Jakub Zaj\k{a}c\textsuperscript{1} and Micha\l{} Dyrek\textsuperscript{1}\par}
\vspace{8pt}
{\sffamily\bfseries\small\textsuperscript{1}AGH University of Krakow\par}

\section*{Abstract}
\noindent\colorbox{abstractbg}{\parbox{\dimexpr\linewidth-2\fboxsep}{\sffamily\small
Portfolio allocation is driven by probabilities of losses, not only by forecasts of the direction (up/down) of
the market. A decision model such as Jev returns a probability for a yes/no question and has been reported to be
well calibrated on a set of benchmarks. However, the model's calibration against realised market outcomes has not
been tested. In this study, every Friday we ask whether the Nasdaq-100 will close lower four weeks later and
compare a two-regime Markov-switching model estimated on data from 1990 to 2026, Jev reading weekly headlines and
geolocated news features computed with spatial queries in PostGIS, and their combination. A knowledge probe shows
that Jev knows events up to late 2024, so we evaluate it on 88 weeks from January 2025 onwards. None of the models
predicts the direction better than the historical base rate. A key finding is that raw Jev overestimates the
probability of a decline: its mean forecast is 0.48 against an observed frequency of 0.38, and its ECE is 0.10. A prior
shift to the historical base rate, which uses no outcomes from the test period, lowers the ECE to 0.015. In
contrast, Platt scaling, isotonic regression and the combination with the Markov model all make the forecasts
worse. In a half-Kelly allocation, the raw Jev strategy returns 3.1\% per year, whereas after the prior shift it
returns 14.5\%; a simple buy-and-hold strategy in QQQ returns 15.3\% per year. The Markov model forecasts large
declines (more than 10\%) well, with an AUC of 0.81, and its strategy has the highest Sharpe ratio over
1999--2026. Asked about such a large decline (base rate 4\%), raw Jev forecasts 16--24\% on average, and the same
prior shift removes this offset. The practical lesson is that decision-model probabilities must be anchored to the base rate of the
market event before they are used to size positions.}}

\vspace{8pt}
\noindent Keywords:\quad calibration, decision models, Markov switching, Kelly criterion, GDELT, PostGIS

\section{Introduction}
\label{sec:intro}

An investor who decides every week how much to put in equities needs more than a forecast of where the market will
go. They also need to know the probability that it will go down. Under the Kelly criterion, both the optimal share
invested and the resulting long-run growth rate of wealth depend on this probability \citep{kelly1956,thorp2006}. A
forecast that ranks weeks well but gives probabilities that are systematically too high or too low leads to
systematic under- or over-investment. Therefore, for allocation it is not enough to discriminate well between
weeks; good calibration, i.e.\ events forecast with probability $p$ occurring with frequency $p$, is just as
important.

The classical way to estimate such probabilities from prices alone is the Markov-switching model
\citep{hamilton1989}, in which returns alternate between a calm bull regime and a volatile bear regime. Regime
probabilities are widely used in asset allocation \citep{ang2002,guidolin2007,nystrup2015}. In addition, there are
so-called \emph{decision models}, e.g.\ Jev, a ``System One'' model that reads any type of input (text or data) and
returns a probability for a yes/no question instead of generating text \citep{typesafe2026jev}. A major feature of
Jev and related models is their calibration, which is reported to be good on a set of common benchmarks
\citep{deusser2026}. These benchmarks consist of labelled questions with a stable base rate. However, when the
frequency of the positive class shifts, the calibration of Jev on financial sentiment deteriorates more than that
of a fine-tuned encoder \citep{iqbal2026}. Whether a decision model is calibrated against \emph{real market
outcomes}, and how much that calibration is worth in an investment decision, has not been tested yet.

We test this for the Nasdaq-100 index. Every Friday we ask the same question, and three sources answer it: a
two-regime Markov-switching model on weekly returns of the index from 1990 to 2026; the Jev decision model, which
reads the most widely published headlines of the week together with features of geolocated news events; and a
combination of the two. We use the GDELT event database \citep{leetaru2013}, which we load into a
PostgreSQL/PostGIS database. On the news side, spatial queries compute weekly conflict and tone measures for seven
regions and for a 200~km radius around major chip fabs and key shipping straits. The same geopolitical features are
also used within the Markov framework through time-varying transition probabilities \citep{diebold1994tvtp}.
Because Jev's knowledge cut-off is not published, we measure it with a knowledge probe and use Jev only after it,
from January 2025 to October 2026 (88 evaluated weeks). Besides the direction, we ask both models about a large
decline of more than 10\%, the rare event in which the Markov model has skill. We compare the calibration of the models with proper scoring
rules, test simple recalibration methods that use only past information, and turn the forecasts into a weekly
half-Kelly allocation between QQQ and T-bills.

Our three research questions are: \textbf{RQ1}: Are Jev's probabilities of a decline better calibrated than those
of the Markov model? \textbf{RQ2}: Do recalibration and the combination of both models improve calibration?
\textbf{RQ3}: Does better calibration lead to better allocation? Our three hypotheses are: H1: Jev is not worse
calibrated than the Markov model; H2: recalibration improves both models; H3: the combination of both models yields
the highest growth rate.

We make three main contributions:
(i)~we test, for the first time, the calibration of a decision model against realised index returns, and we measure
its knowledge cut-off directly;
(ii)~we compare it with, and combine it with, a Markov-switching model, with and without geopolitical transition
probabilities derived from spatial queries on geolocated news;
(iii)~we evaluate the economic value of calibration with a Kelly allocation.
We find that the direction of the Nasdaq-100 over four weeks cannot be predicted by any of the tested models. Raw
Jev overstates the probability of a decline by about ten percentage points, which leaves a Kelly investor almost
out of the market in a rising year. However, a simple one-parameter prior shift that uses no outcomes from the test
period removes this bias. The Markov model stays close to the base rate for the direction but forecasts large
declines well, and its raw probabilities give the best allocation over 1999--2026.

\section{Related Work}
\label{sec:related}

\paragraph{Regimes and allocation.}
\citet{hamilton1989} introduced Markov switching for modelling economic time series, and the filtered regime
probabilities are a natural input for a dynamic allocation strategy. For international allocation, regime switches
in correlations and volatilities have been found to matter \citep{ang2002}. Markov switching has also been used for
multivariate time series to derive optimal portfolios in a dynamic setting \citep{guidolin2007}, and a regime-based
allocation strategy has been found to outperform static portfolios \citep{nystrup2015}. Transition probabilities
can also depend on observed variables \citep{diebold1994tvtp,filardo1994}, which we use for geopolitical features.
In all these studies, however, the focus is on evaluating the portfolios built from the regime probabilities rather
than on the calibration of these probabilities themselves.

\paragraph{Text, news and geopolitical risk.}
\citet{tetlock2007} shows that pessimism in financial news predicts downward pressure on stock prices, followed by
a reversal. Related work includes domain-specific language models for financial sentiment \citep{araci2019} and
large language models that read news headlines to predict stock returns \citep{lopezlira2023}. These studies,
however, evaluate only the direction of the predictions or the returns, not the calibration of the probabilities. A
geopolitical risk index has been built from newspaper text and related to economic activity and downside risks
\citep{caldara2022}. GDELT \citep{leetaru2013} codes news events around the world by location, type, tone and
source, which makes it possible to build geographically targeted risk measures.

\paragraph{Calibration and decision models.}
Proper scoring rules such as the Brier score and the log-loss reward calibrated forecasts
\citep{brier1950,gneiting2007}. Modern neural networks, however, are often miscalibrated \citep{guo2017}, and
simple post-hoc maps such as Platt scaling \citep{platt1999} and isotonic regression \citep{zadrozny2002} are used
as a fix. When only the base rate of the target event changes, a classifier's outputs can be adjusted to the new
prior probabilities \citep{saerens2002}. Decision models are trained to output calibrated probabilities directly
\citep{typesafe2026jev,deusser2026}, but their calibration is known to suffer under prevalence shift
\citep{iqbal2026}. The Kelly criterion links the probabilities used in an investment decision to the growth rate of
wealth \citep{kelly1956,thorp2006}. None of these works, however, evaluates a decision model designed to be
calibrated against realised market outcomes, or measures the value of its calibration in an actual allocation
decision.

\section{Data}
\label{sec:data}

\paragraph{Prices.}
The data consist of daily closing values of the Nasdaq-100 index (\texttt{\^{}NDX}) from October 1985, of the QQQ
ETF (adjusted for distributions) from March 1999, and of the 13-week T-bill rate (\texttt{\^{}IRX}), all from Yahoo
Finance. A week ends on Friday, i.e.\ a week's value is the last close of the week; weeks that have not yet ended
are dropped. In total, there are 2142 weekly log returns from October 1985 to 23 October 2026;
Table~\ref{tab:descriptive} reports descriptive statistics of these returns and the base rates of the two target
events. The index closed lower four weeks later in 36.7\% of all weeks, and in 4.1\% of all weeks it closed more
than 10\% lower four weeks later (in the evaluation samples of the Markov model in Tables~\ref{tab:calibration} and~\ref{tab:robustness}, the
two rates are 36.5\% and 4.5\%).

\paragraph{Geolocated events (GDELT and PostGIS).}
We use the GDELT 2.0 event table \citep{leetaru2013}, which records news reports worldwide with the actors
involved, the type of event (CAMEO code), a Goldstein score of the event's potential impact on the stability of a
country, the average tone of the articles and the location of the action. We select root events with coordinates
added between 1 March 2015 and Friday 23 October 2026, 20:00~UTC. Each event is assigned to the week in which GDELT
recorded it (field \texttt{DATEADDED}), not the week in which it occurred, so that a week contains only information
available at its end. Events added on Friday after 20:00~UTC (i.e.\ after the Nasdaq close) belong to the following
week. BigQuery aggregates the events by week~$\times$~$0.1^\circ$ cell~$\times$~country~$\times$~geocoding type,
with counts and sums of tone, Goldstein scores and mentions: 15.3 million rows representing 552 million events.

The rows are loaded into PostgreSQL with PostGIS as points (SRID 4326) and combined with two additional layers. The
first, \texttt{regions}, contains seven polygons built from Natural Earth country polygons: the United States,
China (including Hong Kong and Macau), Taiwan, South Korea, Japan, Europe (including Russia west of $60^\circ$E)
and the Middle East (the Gulf states, the Levant, Iraq, Iran, Turkey, Egypt, Yemen and Cyprus). The polygons are
buffered by about 15~km so that coastal cells rounded to $0.1^\circ$ remain inside; a cell that falls into the
buffers of more than one region is assigned to the nearest unbuffered region. The second layer,
\texttt{key\_sites}, contains seven chip fabs (TSMC in Hsinchu, Taichung and Tainan; Samsung in Hwaseong and
Pyeongtaek; Foxconn in Zhengzhou and Shenzhen) and five chokepoints (the Taiwan Strait, the Strait of Hormuz, the
Strait of Malacca, Bab el-Mandeb and the Suez Canal). For each week, \texttt{ST\_Within} assigns the events to the
corresponding regions, and \texttt{ST\_DWithin} on the geography type counts all events within 200~km of a key
site. Events geocoded only to a country, rather than to a region or city, are excluded from this count, because
GDELT places them at the centroid of the country, and for Taiwan this centroid lies close to the TSMC fabs. The
output contains the number of events, the number of conflict events (CAMEO codes 18--20), the average tone and the
average Goldstein score per region and week.

GDELT is also affected by outages. Using hourly totals of all events, we drop weeks in which less than 80\% of the
hours are covered. This affects 14 of 608 weeks: seven weeks in October--November 2020, the three weeks from 20 June to 4 July
2025, which fall in the test period, and four single weeks in 2015, 2017, 2021 and 2023. A plausibility check confirms that the features react to known
events. The conflict share in the Middle East in the week of the Hamas attack on Israel (week ending 13 October
2023) is at the 98.8th percentile of all weeks of that region, and in the week of the Iran--Israel strikes of April
2024 (week ending 19 April 2024) at the 98.3rd percentile. In practice, the 200~km count is dominated by news from
Seoul and Taipei, which lie close to the fabs, so we use the conflict share as our main feature.

\paragraph{Headlines.}
Headlines are taken from the titles of articles in the GDELT Global Knowledge Graph. For each week we consider
English-language articles and assign them to economy and markets or to conflict and geopolitics based on keywords
in the article's title. Identical titles are merged and ranked by the number of distinct sources that published
them, which measures how widely a story was syndicated. We exclude sources with \texttt{.uk} and \texttt{.ie}
domains, because syndication by a single British news agency to hundreds of local papers would otherwise dominate
the ranking. The remaining top stories are mostly US news distributed by wire services. We keep titles published by
at least two sources and select the 25 most widely published titles for each of the two topics and each week. In
total, we have 4,700 headlines for 94 weeks from 3 January 2025 to 23 October 2026, with the week of 27 June 2025
missing because of the GDELT outage. As for the events, articles processed by GDELT after Friday 20:00~UTC (i.e.\
after the Nasdaq close) are assigned to the following week.

\paragraph{Map data.}
For the map in the Results, we use the 30 most mentioned conflict events (CAMEO codes 15--20) of each day between
March 2015 and October 2026, 127,110 events in total, each with its location (latitude and longitude).

\begin{table}[tbp]
\centering
\footnotesize
\setlength{\tabcolsep}{4pt}
\begin{adjustbox}{max width=\linewidth}
\begin{tabular}{llrrrrrrr}
\toprule
 & & & \multicolumn{4}{c}{Weekly log return (\%)} & \multicolumn{2}{c}{Base rate, $H=4$} \\
\cmidrule(lr){4-7} \cmidrule(lr){8-9}
Sample & Years & Weeks & Mean & Std. & Skew. & Ex. kurt. & Any decline & Decline $>10\%$ \\
\midrule
Full & 1985--2026 & 2142 & 0.26 & 3.29 & $-$0.77 & 6.44 & 0.367 & 0.041 \\
Evaluation & 1992--2026 & 1771 & 0.26 & 3.34 & $-$0.65 & 5.89 & 0.365 & 0.043 \\
\bottomrule
\end{tabular}
\end{adjustbox}
\caption[Descriptive statistics of the Nasdaq-100.]{Descriptive statistics of the Nasdaq-100. Weeks end on Friday. Base rate: share of weeks after which the index is lower (or more than 10\% lower) 4 weeks later. The evaluation sample contains the weeks for which every model in Table~\ref{tab:calibration} has a forecast and the outcome is known (1992-10-23 to 2026-09-25).}
\label{tab:descriptive}
\end{table}

\section{Methods}
\label{sec:methods}

Each week $t$ ends with the Nasdaq-100 close on Friday. The target is $Y_t = \mathbf{1}\{\sum_{k=1}^{H} r_{t+k} <
0\}$ with $H = 4$ weeks, where $r_t$ is the weekly log return of the Nasdaq-100 in per cent; the target thus
indicates whether the index closes lower four weeks later. The second event, a \emph{large decline}, is
$Y^{10}_t = \mathbf{1}\{\sum_{k=1}^{H} r_{t+k} < 100 \log 0.9\}$, i.e.\ the close $H$ weeks later is more than
10\% below the close at $t$ as a simple return ($-10\%$ corresponds to a sum of log returns of $-10.5\%$). Both
events compare two closes; a fall that is partly recovered before $t+H$ does not count. All forecasts $p_t = \Pr(Y_t = 1)$ are based only on
information available at the close of week $t$. Any model that is fitted to outcomes (for recalibration, for the
combination or for the Kelly moments) uses only pairs $(p_s, Y_s)$ with $s \le t - H$, i.e.\ pairs whose outcome is
already known at $t$.

\subsection{Markov-switching model}
\label{sec:markov}
Weekly returns are modelled by a two-regime model with switching mean and variance \citep{hamilton1989},
\begin{equation}
  r_t = \mu_{S_t} + \sigma_{S_t}\varepsilon_t, \qquad \varepsilon_t \sim \mathcal{N}(0,1), \qquad
  S_t \in \{\text{bull}, \text{bear}\}, \qquad \Pr(S_{t+1} = j \mid S_t = i) = P_{ij},
\end{equation}
where the bear regime is the one with the lower mean. We estimate the model by maximum likelihood on an expanding
window, starting after five years of data (the first forecast is in October 1990), and re-estimate it every 13
weeks. Between re-estimations, the Hamilton filter updates the filtered regime probabilities $\pi_t = \Pr(S_t \mid
r_1, \dots, r_t)$ with fixed parameters. Given $\pi_t$, the forecast is exact and is obtained by summing over all
$2^H$ possible regime paths $s = (s_1, \dots, s_H)$,
\begin{equation}
  p_t^{\mathrm{MS}} = \sum_{s} \Pr(s \mid \pi_t, P)\;
  \Phi\!\left(\frac{-\sum_{k} \mu_{s_k}}{\sqrt{\sum_{k} \sigma^2_{s_k}}}\right).
\end{equation}

\paragraph{Geopolitical transition probabilities.}
In an extension of the model, the transition probabilities depend on GDELT data (see Data) through a logistic link
function \citep{diebold1994tvtp}: $\Pr(S_{t+1} = \text{bull} \mid S_t = i, z_t) = \Lambda(a_i + b_i z_t)$. Here,
$z_t$ is the share of conflict events (CAMEO codes 18--20) among all geolocated events in the seven regions in week
$t$. We use a share rather than a count because the number of articles in GDELT varies over time. The conflict
share $z_t$ is standardised with the mean and standard deviation of past weeks only (at least 52 weeks). Before
GDELT 2.0 starts (March 2015) and during the first year after that, $z_t = 0$, so the transition probabilities are
those of the baseline model. A version with two features (the conflict share and the average tone) gave unstable
estimates for the exit from the rare bear regime, so the main specification has only one feature. For the $H$-week
forecast, the transition matrix is evaluated at $z_t$ for all weeks of the horizon.

\subsection{Decision model (Jev)}
\label{sec:jev}
Jev is a ``System One'' decision model: instead of generating text, it returns a probability for a yes/no question
\citep{typesafe2026jev,deusser2026}. We query version \texttt{typesafe/jev-1.14-20261014} through the OpenRouter
Decisions API with one fixed question for all weeks and inputs: \emph{``Given the information available on [date]
(a Friday), will the Nasdaq-100 index close lower 4 weeks later than it closed on [date]?''}, with the criteria
``closes lower'' (true) and ``closes at the same level or higher'' (false). The output $p_t^{\mathrm{Jev}}$ is the
probability of ``true''. We query Jev with four alternative inputs: (i) the 50 most widely published
English-language headlines of the week (25 on economy and markets, 25 on conflict and geopolitics; see Data); (ii)
the same headlines plus a JSON object with the GDELT features of the seven regions and their percentiles relative
to past weeks; (iii) this JSON object alone; and (iv) a JSON object with Nasdaq-100 returns over 1, 4, 13 and 52
weeks, the 13-week volatility and the drawdown from the 52-week high (no index level, so that the input does not
reveal the date). All 373 queries were made on 27 October 2026 and cost USD~0.017 in total (about USD~0.00005 per
query). For the large decline we ask the same model a second question with inputs (i) and (iv): \emph{``\dots\ will
the Nasdaq-100 index close more than 10\% lower 4 weeks later than it closed on [date]?''} (189 queries on 15
November 2026, USD~0.007; the API returned the same version). All responses are cached for reproducibility.

\paragraph{Knowledge cut-off.}
Jev's training cut-off is not published, so we measure it with a knowledge probe. We ask Jev whether pairs of
statements about market events are true: one statement is true, the other is its false counterpart. If Jev knows a
period, $\Pr(\text{true})$ is much higher for the true statement than for the false one. The gap is 0.88 for 2022
and 0.91 for March 2023, drops to 0.21--0.45 for August to November 2024, and falls to $-0.08$ and $0.01$ for
January and April 2025, respectively. We therefore use Jev only from January 2025 (see the test sample below). The
same probe shows that earlier weeks cannot be used to train recalibration maps without look-ahead
bias.

\paragraph{Test sample.}
The test period has 95 Fridays, from 3 January 2025 to 23 October 2026. The outcome of week $t$ is known at
$t + 4$, so outcomes exist for the 91 weeks up to 25 September 2026. The prior shift (Eq.~\ref{eq:shift}) needs a
mean of at least four Jev forecasts, so the first three weeks (3--17 January 2025) serve only to initialise it, and
all Jev variants are evaluated on the same weeks from 24 January 2025: 88 weeks for the price input, 87 with
headlines (the week of 27 June 2025 has no headlines) and 85 for the inputs with geo features (the GDELT outage
removes the weeks of 20 June, 27 June and 4 July 2025).

\subsection{Recalibration and combination}
\label{sec:recal}
We compare three recalibration maps, all of which are estimated on an expanding window of past pairs with known
outcomes. The three approaches are Platt scaling, a logistic regression of $Y_s$ on $\mathrm{logit}\,p_s$
\citep{platt1999}; isotonic regression \citep{zadrozny2002}; and, for Jev, a simple \emph{prior shift},
\begin{equation}
  \mathrm{logit}\,\tilde p_t = \mathrm{logit}\,p_t \;-\; \frac{1}{t'}\sum_{s \le t} \mathrm{logit}\,p_s
  \;+\; \mathrm{logit}\,\bar y_{\le t-H},
  \label{eq:shift}
\end{equation}
where the mean is taken over the $t'$ Jev forecasts made up to and including week $t$ (forecasts, not outcomes, so
they are known at $t$), and $\bar y_{\le t-H}$ is the historical frequency of $Y = 1$ among the outcomes known at
time $t$, i.e.\ over the weeks $s \le t - H$ since 1990. The prior shift moves the average Jev forecast to the
long-run base rate but keeps the week-to-week variation of the Jev forecasts. Its anchor $\bar y_{\le t-H}$ is
dominated by 35 years of index history, and the shift needs no fitting to outcomes from the test period; it is
available once four Jev forecasts exist (cf.\ the prior adjustment of \citealp{saerens2002}). Platt and isotonic maps, in contrast, start after 104 known pairs for the Markov model and
after 26 pairs for Jev. For Jev, we use L2 regularisation ($C = 1$) in Platt scaling and clip the outputs of
isotonic regression to $[0.01, 0.99]$. The combination is a logistic regression $\mathrm{logit}\,p_t^{\mathrm{C}} =
a + b\,\mathrm{logit}\,p_t^{\mathrm{MS}} + c\,\mathrm{logit}\,p_t^{\mathrm{Jev}}$, estimated in the same way as
Platt scaling for Jev.

\subsection{Evaluation}
\label{sec:eval}
We report a set of standard performance measures: the Brier score $\frac{1}{N}\sum (p_t - Y_t)^2$
\citep{brier1950}, the log-loss $-\frac{1}{N}\sum [Y_t \log p_t + (1 - Y_t)\log(1 - p_t)]$, the expected
calibration error (ECE: ten equal-width bins, mean absolute gap between average forecast and observed frequency,
weighted by bin size) and the AUC. The benchmark is \emph{climatology}, i.e.\ the frequency of $Y = 1$ among the
outcomes known at time $t$. Differences in Brier score and log-loss are tested with the Diebold--Mariano test
\citep{dieboldmariano1995}. Since the forecasts overlap ($H = 4$), we estimate the variance of the mean loss
differential with a Newey--West (Bartlett) kernel with $\max\{H - 1, \lfloor 4 (N/100)^{2/9} \rfloor\}$ lags
\citep{neweywest1987}. Confidence intervals come from a circular block bootstrap \citep{politis1992} with a block
length of 26 weeks and 2000 draws.

\subsection{Allocation}
\label{sec:kelly}
Each Friday, the investor holds a share $w_t$ of QQQ and $1 - w_t$ in T-bills until the next Friday. The share
follows a fractional Kelly rule \citep{kelly1956,thorp2006},
\begin{equation}
  w_t = \mathrm{clip}\!\left(f\,\frac{\hat\mu(p_t) - r_{f,t}}{\hat\sigma^2(p_t)},\, 0,\, 1\right), \qquad f = 0.5,
  \label{eq:kelly}
\end{equation}
where $r_{f,t}$ is the $H$-week T-bill return and $\hat\mu(p_t)$ and $\hat\sigma^2(p_t)$ are the mean and variance
of the $H$-week return of a mixture: with probability $p_t$, the distribution of past returns that ended lower
($Y_s = 1$), and with probability $1 - p_t$, the distribution of the other past returns ($Y_s = 0$). Both
conditional distributions are estimated from known outcomes only, and the rule starts once at least 104 outcomes
are known. The same rule is applied to every forecast $p_t$, so the strategies differ only in the probability they
use. The forecast refers to four weeks, whereas the position is held for one week. We use the $H$-week moments
because they are the moments of the event that $p_t$ forecasts. If returns were independent, mean and variance
would both scale with the horizon and the ratio in Eq.~\eqref{eq:kelly} would not depend on it. However, the
distribution of the next week's return given a four-week decline need not be one quarter of the four-week
distribution. We therefore also compute $\hat\mu(p_t)$ and $\hat\sigma^2(p_t)$ from the \emph{one-week} returns
$r_{s+1}$ of past weeks with $Y_s = 1$ and $Y_s = 0$ (again only for $s \le t - H$) together with the one-week
T-bill rate. This version matches the holding period exactly, and we report it as a robustness check. We also report the simple rule $w_t = 1 - p_t$, buy and hold (100\% QQQ) and a 60/40 portfolio of QQQ and
T-bills rebalanced weekly. Each change in the QQQ share costs 5~bp per unit of turnover, measured against the share
after the previous week's drift. Positions are taken at the Friday close at which the forecast is made.

\begin{table}[tbp]
\centering
\footnotesize
\setlength{\tabcolsep}{4pt}
\begin{adjustbox}{max width=\linewidth}
\begin{tabular}{lccccc}
\toprule
Model & Brier & Log-loss & ECE & AUC & DM \\
\midrule
Climatology & 0.233 & 0.659 & 0.011 & 0.470 & -- \\
 & {\scriptsize [0.224, 0.244]} & {\scriptsize [0.640, 0.681]} & {\scriptsize [0.004, 0.051]} & {\scriptsize [0.419, 0.517]} &  \\
\addlinespace[2pt]
Markov (raw) & 0.232 & 0.658 & 0.031 & 0.496 & $-$0.24 \\
 & {\scriptsize [0.224, 0.241]} & {\scriptsize [0.641, 0.675]} & {\scriptsize [0.011, 0.066]} & {\scriptsize [0.440, 0.547]} & {\scriptsize (0.813)} \\
\addlinespace[2pt]
Markov + Platt & 0.235 & 0.666 & 0.030 & 0.513 & 0.89 \\
 & {\scriptsize [0.224, 0.247]} & {\scriptsize [0.641, 0.693]} & {\scriptsize [0.017, 0.069]} & {\scriptsize [0.459, 0.561]} & {\scriptsize (0.376)} \\
\addlinespace[2pt]
Markov + isotonic & 0.234 & 0.661 & 0.014 & 0.481 & 1.14 \\
 & {\scriptsize [0.224, 0.245]} & {\scriptsize [0.641, 0.683]} & {\scriptsize [0.010, 0.053]} & {\scriptsize [0.429, 0.527]} & {\scriptsize (0.253)} \\
\bottomrule
\end{tabular}
\end{adjustbox}
\caption[Calibration of forecasts of a Nasdaq-100 decline over the next 4 weeks.]{Calibration of forecasts of a Nasdaq-100 decline over the next 4 weeks. 95\% confidence intervals in brackets: circular block bootstrap, blocks of 26 weeks, 2000 draws. DM: Diebold--Mariano statistic on the Brier loss with HAC variance, $p$-value in parentheses; Markov (raw) is compared with climatology, the recalibrated models with Markov (raw). Negative values mean a lower loss than the reference.}
\label{tab:calibration}
\end{table}

\begin{figure}[tbp]
  \centering
  \includegraphics[width=0.85\linewidth]{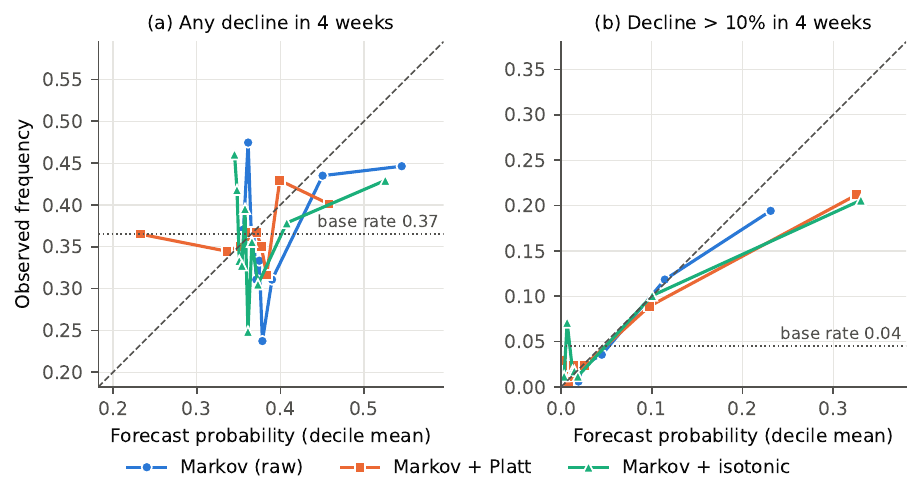}
  \caption{Reliability diagrams of the Markov model for a four-week horizon, 1992--2026. Each point is the mean
    forecast and the observed frequency within one of the ten deciles of the forecast probability. The dashed line
    represents perfect calibration and the dotted line the base rate. (a) Any decline. (b) A decline of more than
    10\%.}
  \label{fig:reliability}
\end{figure}
\begin{table}[tbp]
\centering
\footnotesize
\setlength{\tabcolsep}{4pt}
\begin{adjustbox}{max width=\linewidth}
\begin{tabular}{rlrrrrrrrrr}
\toprule
 & & & & \multicolumn{3}{c}{Brier} & \multicolumn{3}{c}{Log-loss} & \\
\cmidrule(lr){5-7} \cmidrule(lr){8-10}
$H$ & Event & $N$ & Base & Clim. & Markov & DM $p$ & Clim. & Markov & DM $p$ & AUC \\
\midrule
2 & any decline & 1775 & 0.397 & 0.2400 & 0.2399 & 0.913 & 0.673 & 0.673 & 0.906 & 0.501 \\
2 & decline $>10\%$ & 1467 & 0.022 & 0.0215 & 0.0207 & 0.448 & 0.110 & 0.090 & 0.025 & 0.825 \\
4 & any decline & 1771 & 0.365 & 0.2330 & 0.2325 & 0.813 & 0.659 & 0.658 & 0.793 & 0.496 \\
4 & decline $>10\%$ & 1694 & 0.045 & 0.0432 & 0.0392 & 0.045 & 0.191 & 0.152 & 0.006 & 0.808 \\
13 & any decline & 1753 & 0.302 & 0.2140 & 0.2118 & 0.672 & 0.620 & 0.615 & 0.613 & 0.502 \\
13 & decline $>10\%$ & 1753 & 0.100 & 0.0917 & 0.0850 & 0.196 & 0.336 & 0.301 & 0.066 & 0.675 \\
\bottomrule
\end{tabular}
\end{adjustbox}
\caption[Robustness: forecast horizon $H$ (weeks) and definition of the bear event.]{Robustness: forecast horizon $H$ (weeks) and definition of the bear event. Base: base rate of the event. Clim.: climatology. DM $p$: two-sided $p$-value of the Diebold--Mariano test of Markov (raw) against climatology with HAC variance. AUC of Markov (raw).}
\label{tab:robustness}
\end{table}

\begin{figure}[tbp]
  \centering
  \includegraphics[width=0.85\linewidth]{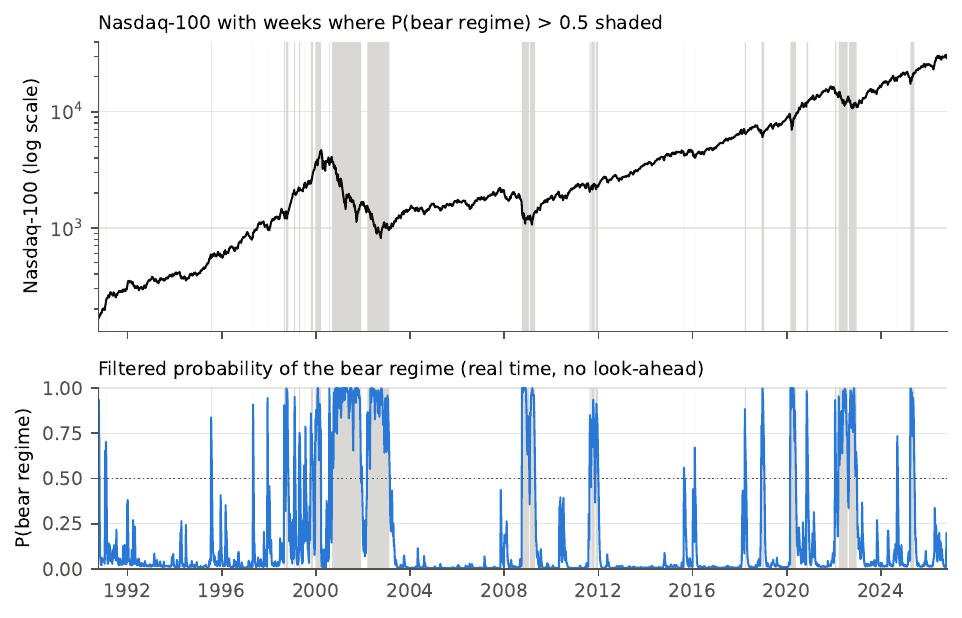}
  \caption{Nasdaq-100 (log scale) and the filtered probability of the bear regime, computed in real time from
    data up to each week. Weeks in which the bear probability exceeds 0.5 are shaded.}
  \label{fig:bear}
\end{figure}
\begin{table}[tbp]
\centering
\footnotesize
\setlength{\tabcolsep}{4pt}
\begin{adjustbox}{max width=\linewidth}
\begin{tabular}{lrrrrrrr}
\toprule
Model & $N$ & Mean $p$ & Brier & Log-loss & ECE & AUC & DM $p$ \\
\midrule
Climatology & 88 & 0.365 & 0.235 & 0.663 & 0.010 & 0.352 & -- \\
Markov (raw) & 88 & 0.381 & 0.251 & 0.696 & 0.129 & 0.323 & 0.049 \\
\addlinespace[2pt]
Jev: headlines & 87 & 0.480 & 0.247 & 0.687 & 0.101 & 0.457 & 0.546 \\
\quad + prior shift & 87 & 0.364 & 0.236 & 0.665 & 0.015 & 0.511 & 0.965 \\
Jev: headlines + geo & 85 & 0.475 & 0.246 & 0.685 & 0.098 & 0.450 & 0.549 \\
\quad + prior shift & 85 & 0.365 & 0.235 & 0.664 & 0.012 & 0.524 & 0.906 \\
Jev: geo features & 85 & 0.453 & 0.240 & 0.673 & 0.077 & 0.555 & 0.734 \\
\quad + prior shift & 85 & 0.367 & 0.235 & 0.662 & 0.009 & 0.489 & 0.260 \\
Jev: prices only & 88 & 0.466 & 0.243 & 0.678 & 0.091 & 0.515 & 0.621 \\
\quad + prior shift & 88 & 0.364 & 0.234 & 0.660 & 0.011 & 0.549 & 0.646 \\
\bottomrule
\end{tabular}
\end{adjustbox}
\caption[Jev and the Markov model in the test period (2025-01-24 to 2026-09-25), decline over the next 4 weeks.]{Jev and the Markov model in the test period (2025-01-24 to 2026-09-25), decline over the next 4 weeks. Base rate of the event: 0.375. Jev version: typesafe/jev-1.14-20261014. Prior shift: $\mathrm{logit}\,\tilde p_t = \mathrm{logit}\,p_t - \overline{\mathrm{logit}\,p}_{\le t} + \mathrm{logit}\,\bar y_{\le t-H}$, i.e. the mean of past Jev forecasts is moved to the historical base rate (since 1990); it uses no outcomes from the test period. $N$ differs by up to 3 weeks because GDELT coverage gaps remove some weeks from the geo variants. DM $p$: two-sided $p$-value of the Diebold--Mariano test on the Brier loss against climatology (HAC variance).}
\label{tab:jev}
\end{table}

\begin{figure}[tbp]
  \centering
  \includegraphics[width=0.85\linewidth]{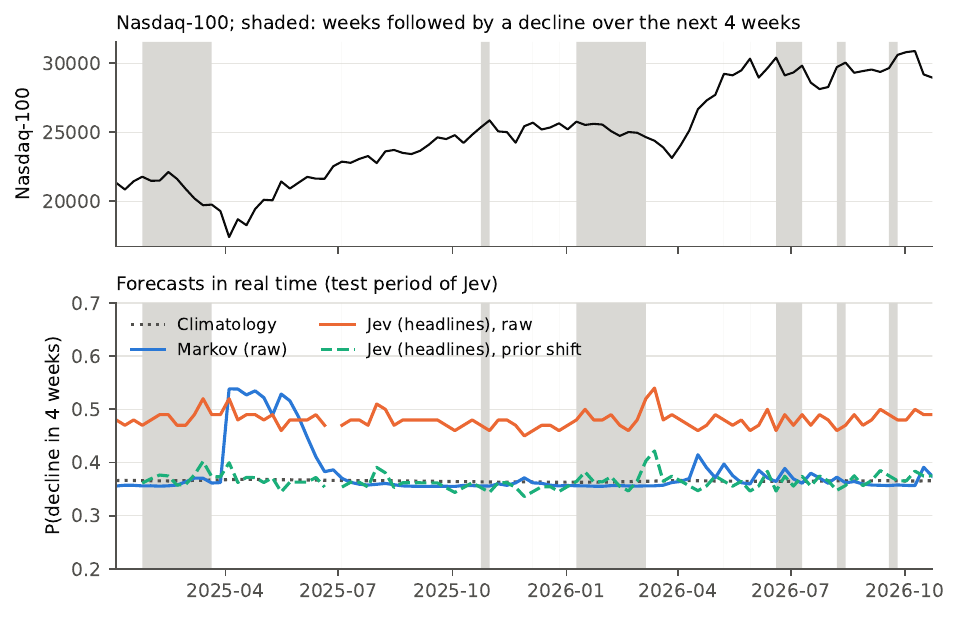}
  \caption{Test period of Jev, January 2025 to October 2026. Top: the Nasdaq-100; shaded weeks are those
    followed by a decline over the next four weeks. Bottom: forecasts of a decline from climatology, the Markov
    model and Jev with headlines, both raw and after the prior shift. The gap in June 2025 is due to a GDELT
    outage.}
  \label{fig:jev}
\end{figure}
\begin{table}[tbp]
\centering
\footnotesize
\setlength{\tabcolsep}{4pt}
\begin{adjustbox}{max width=\linewidth}
\begin{tabular}{lrrrcrrrr}
\toprule
Strategy & CAGR & Vol. & Sharpe & $\Delta$Sharpe & Max DD & Log gr. & Avg. $w$ & Turn. \\
\midrule
Buy and hold (100\% QQQ) & 10.7 & 24.4 & 0.46 & -- & $-$82.7 & 10.2 & 1.00 & 0.04 \\
60/40 (QQQ / T-bill) & 7.9 & 14.6 & 0.46 & 0.00 & $-$61.2 & 7.6 & 0.60 & 0.33 \\
 &  &  &  & {\scriptsize [0.00, 0.00]} &  &  &  &  \\
Half-Kelly: climatology & 10.5 & 24.2 & 0.45 & $-$0.01 & $-$82.6 & 10.0 & 0.99 & 0.08 \\
 &  &  &  & {\scriptsize [$-$0.01, 0.00]} &  &  &  &  \\
Half-Kelly: Markov (raw) & 12.2 & 18.9 & 0.60 & $+$0.14 & $-$55.6 & 11.5 & 0.83 & 1.85 \\
 &  &  &  & {\scriptsize [$-$0.07, $+$0.38]} &  &  &  &  \\
Half-Kelly: Markov + Platt & 9.5 & 21.7 & 0.44 & $-$0.02 & $-$79.9 & 9.1 & 0.92 & 1.10 \\
 &  &  &  & {\scriptsize [$-$0.12, $+$0.09]} &  &  &  &  \\
Half-Kelly: Markov + isotonic & 10.3 & 19.9 & 0.49 & $+$0.03 & $-$69.6 & 9.8 & 0.87 & 1.84 \\
 &  &  &  & {\scriptsize [$-$0.13, $+$0.22]} &  &  &  &  \\
$w = 1 - p$: Markov (raw) & 8.3 & 13.7 & 0.51 & $+$0.05 & $-$55.8 & 8.0 & 0.60 & 0.57 \\
 &  &  &  & {\scriptsize [0.00, $+$0.10]} &  &  &  &  \\
\bottomrule
\end{tabular}
\end{adjustbox}
\caption[Weekly allocation between QQQ and T-bills, 1999-03-12 to 2026-10-16.]{Weekly allocation between QQQ and T-bills, 1999-03-12 to 2026-10-16. Half-Kelly: $w_t = \mathrm{clip}(f(\hat\mu(p_t) - r_f)/\hat\sigma^2(p_t), 0, 1)$ with $f = 0.5$. CAGR, volatility (Vol.), maximum drawdown (Max DD) and log growth (Log gr.) in \% per year. Avg. $w$: mean share of QQQ. Turn.: yearly sum of $|\Delta w|$. Transaction costs of 5 bp per unit of turnover. $\Delta$Sharpe: difference to buy and hold with a 95\% block-bootstrap confidence interval.}
\label{tab:allocation}
\end{table}

\begin{table}[tbp]
\centering
\footnotesize
\setlength{\tabcolsep}{4pt}
\begin{adjustbox}{max width=\linewidth}
\begin{tabular}{lrrrrrrrrr}
\toprule
 & \multicolumn{3}{c}{Log growth} & \multicolumn{3}{c}{Sharpe} & \multicolumn{3}{c}{Max DD} \\
\cmidrule(lr){2-4} \cmidrule(lr){5-7} \cmidrule(lr){8-10}
Kelly with & $f=0.25$ & $f=0.5$ & $f=1$ & $f=0.25$ & $f=0.5$ & $f=1$ & $f=0.25$ & $f=0.5$ & $f=1$ \\
\midrule
climatology & 6.7 & 10.0 & 10.2 & 0.36 & 0.45 & 0.46 & $-$74.9 & $-$82.6 & $-$82.7 \\
Markov (raw) & 8.7 & 11.5 & 11.9 & 0.55 & 0.60 & 0.61 & $-$45.0 & $-$55.6 & $-$56.7 \\
Markov + Platt & 6.4 & 9.1 & 8.8 & 0.35 & 0.44 & 0.41 & $-$74.0 & $-$79.9 & $-$84.3 \\
Markov + isotonic & 7.9 & 9.8 & 9.6 & 0.45 & 0.49 & 0.47 & $-$62.1 & $-$69.6 & $-$74.8 \\
\bottomrule
\end{tabular}
\end{adjustbox}
\caption[Robustness: Kelly fraction $f$.]{Robustness: Kelly fraction $f$. Log growth and maximum drawdown in \% per year; costs as in Table~\ref{tab:allocation}. With climatology and $f=1$ the weight is always 1 (buy and hold).}
\label{tab:kelly_f}
\end{table}

\begin{table}[tbp]
\centering
\footnotesize
\setlength{\tabcolsep}{4pt}
\begin{adjustbox}{max width=\linewidth}
\begin{tabular}{lrrrrrrr}
\toprule
Strategy & CAGR & Vol. & Sharpe & Max DD & Log gr. & Avg. $w$ & Turn. \\
\midrule
Buy and hold (100\% QQQ) & 15.3 & 20.7 & 0.61 & $-$21.3 & 14.3 & 1.00 & 0.59 \\
60/40 (QQQ / T-bill) & 11.2 & 12.4 & 0.61 & $-$13.1 & 10.6 & 0.60 & 0.63 \\
Half-Kelly: Markov (raw) & 1.0 & 18.1 & $-$0.07 & $-$21.3 & 1.0 & 0.87 & 2.58 \\
Half-Kelly: Jev (headlines) & 3.1 & 0.7 & $-$1.26 & $-$0.8 & 3.0 & 0.01 & 1.09 \\
\quad + prior shift & 14.5 & 20.2 & 0.58 & $-$21.4 & 13.5 & 0.98 & 2.07 \\
Half-Kelly: Jev (headlines + geo) & 3.3 & 0.9 & $-$0.64 & $-$1.0 & 3.3 & 0.02 & 1.95 \\
\quad + prior shift & 15.1 & 20.2 & 0.61 & $-$21.1 & 14.1 & 0.98 & 1.87 \\
Half-Kelly: Jev (geo features) & 7.1 & 3.3 & 0.91 & $-$2.7 & 6.8 & 0.14 & 3.39 \\
\quad + prior shift & 15.3 & 20.7 & 0.61 & $-$21.3 & 14.3 & 1.00 & 0.59 \\
Half-Kelly: Jev (prices only) & 5.5 & 2.7 & 0.57 & $-$1.0 & 5.4 & 0.09 & 3.88 \\
\quad + prior shift & 16.3 & 19.7 & 0.67 & $-$19.1 & 15.1 & 0.98 & 1.97 \\
\bottomrule
\end{tabular}
\end{adjustbox}
\caption[Allocation in the test period, 2025-01-24 to 2026-10-16.]{Allocation in the test period, 2025-01-24 to 2026-10-16. Same rules and costs as in Table~\ref{tab:allocation}, $f = 0.5$. Half-Kelly with climatology holds 100\% QQQ in this period (identical to buy and hold). Only 88 weeks: the numbers illustrate the effect of calibration, they are not a test.}
\label{tab:allocation_jev}
\end{table}

\begin{figure}[tbp]
  \centering
  \includegraphics[width=0.85\linewidth]{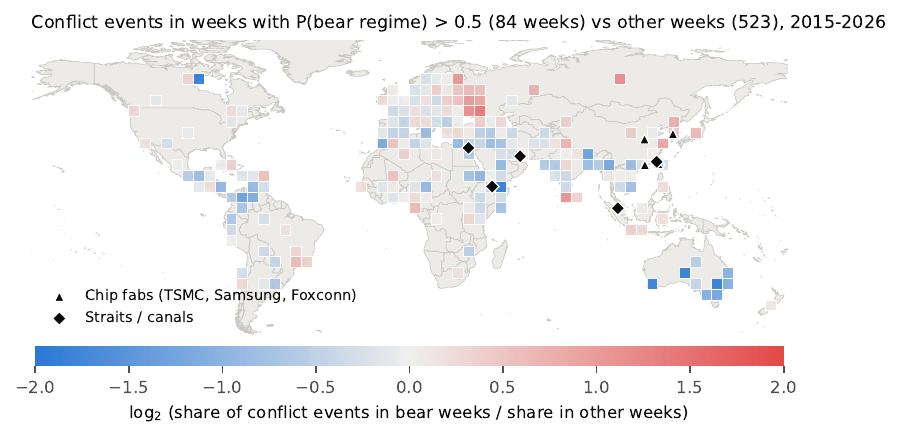}
  \caption{Location of conflict events (GDELT, 30 most mentioned events per day) in weeks in which the bear
    probability is above 0.5, compared with all other weeks, 2015--2026. Only cells of $5^\circ \times 5^\circ$
    with at least 100 events are shown; the colour of each cell is $\log_2$ of the ratio of the cell's share of
    events in bear weeks to its share in the other weeks. Markers show the chip fabs and straits used for the
    200~km features.}
  \label{fig:map}
\end{figure}

\section{Results}
\label{sec:results}

\subsection{Calibration of the Markov model}
\label{sec:res-markov}
Table~\ref{tab:calibration} reports the out-of-sample evaluation over 1771 weeks (October 1992 to September 2026).
The Markov model adds almost nothing to climatology: its Brier score is 0.232 against 0.233 (DM $p = 0.81$), and
its AUC of 0.50 shows that it does not separate weeks followed by a decline from the others. Its forecasts are also
very narrow: half of them lie between 0.36 and 0.39, and all of them lie between 0.35 and 0.60. In the reliability
diagram (Figure~\ref{fig:reliability}a), the observed frequency in the deciles of $p_t^{\mathrm{MS}}$ scatters
around the base rate of 0.37 without following the diagonal. Recalibration does not improve the results for this
event either: Platt scaling and isotonic regression raise the Brier score slightly (DM $p = 0.38$ and $0.25$).
Isotonic regression lowers the ECE from 0.031 to 0.014 only by moving the forecasts towards the base rate.

The picture changes when the event is a large decline, i.e.\ a close more than 10\% below the current one four weeks later (base rate
4.5\%). Here the Markov model beats climatology on both losses: the Brier score falls to 0.039 against 0.043 (DM $p
= 0.045$) and the log-loss to 0.152 against 0.191 ($p = 0.006$), while the AUC for this event is 0.81
(Table~\ref{tab:robustness}). Its forecasts are close to the diagonal (Figure~\ref{fig:reliability}b), whereas the
recalibrated versions overstate the probability in the top decile. The same holds for a two-week horizon (AUC 0.83,
log-loss $p = 0.025$). At 13 weeks the advantage shrinks (AUC 0.67, log-loss $p = 0.066$), and for any decline no
horizon gives a significant improvement. In other words, the model captures periods of high risk but does not
predict the direction of the market. On the full sample, the bear regime has a weekly mean return of $-0.39\%$ and
a volatility of 5.5\%, whereas the bull regime has a weekly mean return of $+0.45\%$ and a volatility of 2.3\%. The
expected duration is about 20 weeks for the bear regime and about 70 weeks for the bull regime.
Figure~\ref{fig:bear} shows that the filtered bear probability exceeds 0.5 mainly in 1998--2003, 2008--2009, 2011, 2020
and 2022, and for shorter spells in 2018 and 2025.

\subsection{Geopolitical transition probabilities}
\label{sec:res-tvtp}
Letting the transition probabilities depend on the GDELT conflict share improves the baseline Markov model for the
direction: over 497 weeks from March 2017, the Brier score falls from 0.2267 to 0.2219 (DM $p = 0.014$; log-loss
0.646 vs 0.636, $p = 0.013$). In this period, however, the baseline is slightly worse than climatology (Brier
0.2233), and the geo model does not beat climatology (DM $p = 0.73$), so the direction remains unpredictable; the
conflict share only removes part of the baseline's deficit. For a
decline of more than 10\%, the geo model is better than the baseline Markov model (log-loss 0.140 vs 0.149, DM $p =
0.025$; Brier $p = 0.067$; AUC 0.75 vs 0.71; 414 weeks from October 2018). In the last estimation, a higher
conflict share lowers the probability of leaving the bear regime ($b = -0.72$, s.e.\ 0.44, $p = 0.10$):
geopolitical tension tends to prolong bear regimes rather than to start them. Because this is one of several
specifications we examined, we treat the result only as suggestive.

\subsection{Jev in the test period}
\label{sec:res-jev}
Table~\ref{tab:jev} compares Jev with climatology and the Markov model over the 85--88 weeks from January 2025 for
which the outcome is known. Raw Jev forecasts an average of 0.45--0.48 for every input, while the index closed
lower four weeks later in only 37.5--37.9\% of the weeks. Most of its ECE (0.08--0.10) comes from this constant offset.
Its forecasts also hardly move: with headlines they range only from 0.45 to 0.54, even in the week of the April
2025 tariff announcement (Figure~\ref{fig:jev}). No input beats climatology (DM $p \ge 0.54$), and the area under
the ROC curve (AUC) lies between 0.45 and 0.56 for all inputs. Headlines thus add no information compared with
prices alone.

The prior shift removes the miscalibration: it reduces the ECE to 0.009--0.015 and brings the Brier score to the
level of climatology (0.234--0.236 vs 0.235); the remaining differences are not significant. Platt scaling and
isotonic regression, which need outcomes from the test period and learn from only 26 to 87 past pairs, make the
forecasts worse than climatology (Brier 0.246--0.280 vs 0.234--0.239 in the common sample from July 2025).
Combining Jev with the Markov model does not help either (Brier 0.248--0.251). In the same period, the Markov model
is even worse than climatology (Brier 0.251 vs 0.235, DM $p = 0.049$): after the April 2025 crash, its forecast
jumped to 0.54 just before the market recovered (Figure~\ref{fig:jev}).

\paragraph{Large declines.}
When asked about a decline of more than 10\%, raw Jev forecasts on average 0.24 with headlines (range 0.16--0.37)
and 0.16 with prices only (0.11--0.27), whereas the historical base rate is 0.042. Only 2 of the 87 common test
weeks (14 February and 7 March 2025, both in the sell-off before the April tariff crash) were followed by such a
decline. For this rare event the offset is therefore much larger than for the direction: raw Jev states four to six
times the base rate and is clearly worse than climatology (Brier 0.071 and 0.042 vs 0.023, DM $p < 0.001$; ECE 0.22
and 0.14). After the prior shift, the mean forecast is 0.047 and 0.042, the ECE 0.024 and 0.019, and the Brier score
equals that of climatology (0.0225 and 0.0226 vs 0.0228, DM $p \ge 0.49$). The Markov model, with a mean of 0.036,
is not better than climatology in this period either (Brier 0.026, $p = 0.13$). With two positive weeks from one
episode, the AUC (0.45--0.86 across models) carries no information, so the ability to discriminate large declines
cannot be evaluated in 2025--2026. What the test does show is the prevalence-shift pattern: the rarer the event,
the larger raw Jev's overestimation, and the prior shift removes it for both events.

\subsection{Allocation}
\label{sec:res-kelly}
Over the full period (March 1999 to October 2026, Table~\ref{tab:allocation}), half-Kelly with the raw Markov
forecasts earns 12.2\% per year against 10.7\% for buy and hold, with a Sharpe ratio of 0.60 instead of 0.46 and a
maximum drawdown of $-56\%$ instead of $-83\%$. However, the difference in Sharpe ratio has a wide confidence
interval ($+0.14$, 95\% CI $[-0.07, +0.38]$; the bootstrap share of draws at or below zero is 0.11). Hence the gain
is not statistically established. The recalibrated Markov forecasts do worse than the raw ones (Sharpe ratio 0.44
with Platt scaling and 0.49 with isotonic regression), even though isotonic regression has the lower ECE.
Half-Kelly with climatology almost always holds 100\% QQQ and therefore performs about as well as buy and hold. The
simple rule $w_t = 1 - p_t^{\mathrm{MS}}$ improves the Sharpe ratio by 0.05 (95\% CI $[0.00, +0.10]$), but it lowers the mean equity share to 0.60 and lowers the growth rate. Table~\ref{tab:kelly_f} reports the same ordering for $f =
0.25$ and $f = 1$. Taking the Kelly moments from one-week instead of four-week returns (see the Allocation part of the Methods)
changes almost nothing. For the raw Markov forecasts, the CAGR stays at 12.2\%, the Sharpe ratio is 0.60 in both
versions and the mean QQQ share is 0.83. Over 1999--2026 (Table~\ref{tab:allocation}), the CAGR of every Kelly
strategy changes by at most 0.1 percentage points. In 2025--2026 (Table~\ref{tab:allocation_jev}), it changes by
at most 0.1 points for the Markov model and for Jev after the prior shift, and by at most 0.6 points for raw Jev,
which holds little QQQ in both versions (e.g.\ 14\% vs 21\% on average with geo features).

The test period illustrates what miscalibration costs (Table~\ref{tab:allocation_jev}). With raw Jev forecasts, the
Kelly rule sees an expected return close to the T-bill rate, so it holds on average only 1--14\% QQQ and earns
3--7\% per year, while buy and hold of QQQ earns 15.3\%. After the prior shift, the same forecasts lead to a QQQ
share of 98--100\% and a return of 14.5--16.3\% per year. The Markov strategy earns only 1.0\% in this period,
because it sold after the April 2025 crash. With only 88 weeks of data, these numbers are an illustration, not a
test.

\subsection{Where conflict events concentrate}
\label{sec:res-map}
Figure~\ref{fig:map} compares where conflict events were located in the 84 weeks with a bear probability above 0.5
and in the remaining 523 weeks since 2015. Events around the Black Sea (Russia and Ukraine) are clearly
over-represented in bear weeks (log$_2$ ratio of shares up to 1.5, i.e.\ almost three times their usual share).
Korea and Japan are also over-represented, but to a lesser extent, while Latin America and Australia are
under-represented. Most bear weeks cluster in a few episodes (2020, 2022, 2025), so the map mainly shows which
conflicts dominated the news during these episodes, above all the invasion of Ukraine in 2022. It does not show
that conflicts in these places cause bear markets.

\subsection{Summary}
\label{sec:res-summary}
\textbf{RQ1.} For the direction of the Nasdaq-100 over four weeks, neither model is calibrated in a useful sense.
The Markov model reduces to the base rate, and Jev adds a constant upward offset of about ten percentage points
above the observed frequency of declines (so H1 holds only in the sense that Jev is not worse than the Markov model
in the test period). \textbf{RQ2.} Standard recalibration does not improve either model. A one-parameter prior
shift that uses no outcomes from the test period does, however, correct Jev's offset (so H2 holds only for the
prior shift). The combination of both models does not improve the forecasts, so H3 is not supported. \textbf{RQ3.}
Calibration matters for allocation when it is poor: the offset of raw Jev's forecasts from the base rate keeps a
Kelly investor almost out of the market in a rising year. However, a lower ECE does not by itself translate into a
higher growth rate: the best strategy over the full period 1999--2026 uses the raw Markov forecasts, whose value
lies in detecting periods of high risk.

\section{Discussion and Limitations}
\label{sec:discussion}

\paragraph{What calibration means on markets.}
For the direction of the Nasdaq-100 over four weeks, none of the models has measurable skill, so a well-calibrated
forecast cannot do better than simple climatology. In this setting, the only requirement for a well-calibrated
forecast is to state the correct base rate of a decline, and raw Jev fails exactly at that: it treats a decline
almost as a coin flip, with an average forecast of 0.45--0.48, while declines occurred in only 37.5--37.9\% of the
weeks. The offset is consistent with prevalence shift \citep{iqbal2026}: a model whose answers are calibrated on
benchmark questions with balanced classes has no reason to know the base rate of a market event. The remedy is
correspondingly simple: the prior shift uses only the long history of the index, not Jev's outcomes, and is
available from the fourth week. This is much simpler than standard recalibration, which has to learn a map from
forecasts to outcomes and is therefore the wrong tool when only a few dozen non-overlapping outcomes exist. More
generally, a decision model's calibration on its own benchmarks does not carry over to a domain with a different
base rate, so users should anchor the model to that base rate before acting on its forecasts. The question about
large declines supports this reading: there the base rate is 4\%, raw Jev states 16--24\%, and the same prior
shift brings it back to the level of climatology.

\paragraph{Calibration and economic value.}
However, a lower calibration error is neither necessary nor sufficient for a better allocation. For the Markov
model, isotonic regression lowers the ECE but worsens the portfolio, while the raw Markov forecasts, whose ECE is
higher, give the highest Sharpe ratio and the smallest drawdown over 1999--2026. Their value lies in the
large-decline events that the bear regime detects (AUC 0.81 for a fall of more than 10\%), which a Kelly investor
uses to reduce exposure in turbulent periods. Calibration matters most when it is poor in a systematic direction:
Jev's offset alone costs about eleven percentage points of return in 2025--2026. Geolocated news brings the Markov model's direction forecasts only up to the level
of climatology, but it improves the forecast of large declines, consistent with bear regimes lasting longer when
the conflict share is high.

\paragraph{Role of the spatial pipeline.}
The geographic part of the study yields mostly negative results, and we report them as such. The 200~km features
around fabs and straits tested a specific hypothesis, namely that news close to the semiconductor supply chain and
to shipping chokepoints matters more for a technology index than news elsewhere. The data do not support it: near
the fabs the count is dominated by routine news from Seoul and Taipei, and these features add nothing to the
models. PostGIS is nevertheless needed for the features that we do use. Assigning 15 million weekly cell rows to seven
regions with buffered coastlines is a point-in-polygon problem, and the regional conflict share is the only GDELT
variable with a measurable effect (in the transition probabilities: against the baseline Markov model, DM $p =
0.014$ for the direction and log-loss $p = 0.025$ for large declines). The same
database also supplies the geo inputs given to Jev and the map, which shows that bear weeks coincide with a few
dominant conflicts rather than with a stable spatial pattern. A reader interested only in the main result (the
miscalibration of Jev and its correction) can skip the geographic features; they show that targeted geopolitical news does not rescue the forecasts of the market's direction.

\paragraph{Limitations.}
First, the test period for Jev is short: 88 weeks with known outcomes and overlapping four-week windows, i.e.\
about 22 independent outcomes. The Diebold--Mariano tests and the block bootstrap account for the overlap, but
their power is low, and the allocation results for 2025--2026 are only illustrations. Second, the sample cannot be
extended backwards: the knowledge probe shows that Jev knows events up to at least November 2024, so earlier weeks
would introduce look-ahead bias, both in the evaluation of Jev and in training the recalibration maps. Third, Jev
is a closed model whose versions change; all results refer to \texttt{typesafe/jev-1.14-20261014}, and we cache
every request and response. Fourth, the news input depends on GDELT's coverage: several large wire services and
paywalled outlets are barely represented, the keyword-based topics contain some noise (e.g.\ entertainment stories
that mention large technology companies), and coverage gaps remove some weeks. Fifth, the large-decline results
come from one of several horizons and event definitions that we examined, and the map reflects which conflicts
dominated the news in a few bear episodes rather than a causal effect. Finally, the study covers only one index,
one question and one decision model, and positions are taken at the Friday close at which the forecast is made.

\paragraph{Future work.}
A pre-registered live test would remove any doubt about the knowledge cut-off and would enlarge the sample over
time. It is also the only way to judge whether Jev can discriminate large declines, of which the test period
contains a single episode. Open decision models with a known training cut-off would allow long backtests, and other indices, horizons
and questions (e.g.\ large declines or volatility) would show where decision models add value on markets.

\section{Conclusion}
\label{sec:conclusion}

We tested whether Jev, a decision model advertised as calibrated, gives calibrated probabilities of a decline of
the Nasdaq-100 over four weeks, and compared it with a Markov-switching model, with and without geolocated news
processed in PostGIS. None of the models predicts the direction of the index over this horizon. Jev overestimates
the probability of a decline by about ten percentage points (0.48 on average against an observed frequency of 0.38). A simple
prior shift to the historical base rate reduces its ECE from 0.10 to 0.015 and turns a half-Kelly strategy that
earned 3.1\% per year into one that earns 14.5\%, close to buy and hold. For investors, the lesson is to anchor
decision-model probabilities to the base rate of the market event before sizing positions, and to value models by
the risk episodes they detect rather than by calibration error alone.

\paragraph{Data and code availability.}
The code, the weekly GDELT features, the metadata of the headlines and all Jev requests and responses are
available from the authors. The event data can be downloaded from The GDELT Project
(\url{https://www.gdeltproject.org/}).

\bibliographystyle{apalike}
\bibliography{references}

\end{document}